\documentclass[conference]{IEEEtran}
\IEEEoverridecommandlockouts

\usepackage[normalem]{ulem}
\usepackage[utf8]{inputenc}
\usepackage{amsmath}  
\usepackage{graphicx}
\usepackage{comment}
\usepackage{multirow}
\usepackage{enumitem}
\usepackage{array}
\usepackage{tcolorbox}
\usepackage{xcolor}
\usepackage{balance}
\usepackage{mathtools}
\usepackage{xspace}
\usepackage{url}
\usepackage{eucal}
\usepackage{amsfonts}
\usepackage{color}
\usepackage{booktabs}
\usepackage{bbding}
\usepackage{float}

\usepackage{graphicx}
\usepackage{hyperref}
\usepackage{url}
\usepackage{multirow}
\usepackage{comment}
\usepackage{colortbl}
\usepackage{graphicx, subfigure}

\usepackage{xcolor}
\usepackage{enumitem}
\usepackage{amsmath,amssymb}

\def\BibTeX{{\rm B\kern-.05em{\sc i\kern-.025em b}\kern-.08em
    T\kern-.1667em\lower.7ex\hbox{E}\kern-.125emX}}

\begin{document}
\title{CCBERT: Self-Supervised Code Change Representation Learning}

\makeatletter
\newcommand{\linebreakand}{%
  \end{@IEEEauthorhalign}
  \hfill\mbox{}\par
  \mbox{}\hfill\begin{@IEEEauthorhalign}
}
\makeatother

\author{
    \IEEEauthorblockN{Xin Zhou\IEEEauthorrefmark{2}, Bowen Xu\textsuperscript{*}\thanks{*Corresponding author. Email: bxu22@ncsu.edu}\IEEEauthorrefmark{2}\IEEEauthorrefmark{3}, DongGyun Han\IEEEauthorrefmark{4}, Zhou Yang\IEEEauthorrefmark{2}, Junda He\IEEEauthorrefmark{2}, David Lo\IEEEauthorrefmark{2}}
    \IEEEauthorblockA{\IEEEauthorrefmark{2}\textit{Singapore Management University, Singapore}
    \\\{xinzhou.2020, bowenxu.2017\}@phdcs.smu.edu.sg, \{zyang,jundahe,davidlo\}@smu.edu.sg}
    \IEEEauthorblockA{\IEEEauthorrefmark{3}\textit{North Carolina State University, USA}
    \\bxu22@ncsu.edu}
    \IEEEauthorblockA{\IEEEauthorrefmark{4}\textit{Royal Holloway, University of London, UK}
    \\donggyun.han@rhul.ac.uk}
}

\maketitle

\begin{abstract}

Numerous code changes are made by developers in their daily work, and a superior representation of code changes is desired for effective code change analysis. Recently, Hoang et al. proposed CC2Vec, a neural network-based approach that learns a distributed representation of code changes to capture the semantic intent of the changes. Despite demonstrated effectiveness in multiple tasks, CC2Vec has several limitations: 1) it considers only coarse-grained information about code changes, and 2) it relies on log messages rather than the self-contained content of the code changes.
In this work, we propose CCBERT (\underline{C}ode \underline{C}hange \underline{BERT}), a new Transformer-based pre-trained model that learns a generic representation of code changes based on a large-scale dataset containing massive unlabeled code changes. 
CCBERT is pre-trained on four proposed self-supervised objectives that are specialized for learning code change representations based on the contents of code changes.
CCBERT perceives fine-grained code changes at the token level by learning from the old and new versions of the content, along with the edit actions.
Our experiments demonstrate that CCBERT significantly outperforms CC2Vec or the state-of-the-art approaches of the downstream tasks by 7.7\%--14.0\% in terms of different metrics and tasks. CCBERT consistently outperforms large pre-trained code models, such as CodeBERT, while requiring 6--10$\times$ less training time, 5--30$\times$ less inference time, and 7.9$\times$ less GPU memory.

\end{abstract}

\section{Introduction}
\label{sec:intro}

Software evolves rapidly with the continuous code changes contributed by developers. A strong understanding of code changes, therefore, is vital for the daily development process and maintenance tasks. However, it is also non-trivial since code changes are usually mixed to serve various purposes such as bug fixing, refactoring, and adding new features.
To help developers handle numerous code changes efficiently, many automated solutions have been proposed to address concrete tasks (e.g., defect prediction~\cite{DeepJIT}, patch assessment~\cite{tian2020evaluating}, and bug-fixing commit prediction~\cite{patchnet}).

The initial step for these solutions is to extract a suitable representation of code changes for further processing. Intuitively, following the ``garbage in, garbage out'' principle~\cite{mellin1957work}, the quality of a code change representation plays a major role in determining the eventual outcomes. Such tasks demand a suitable input that boosts the performance of downstream tasks and is computationally convenient to deal with, which motivates our work in developing a superior representation of code changes.

The closest work is by Hoang et al.~\cite{cc2vec}, who proposed a deep neural network model named CC2Vec to learn the distributed representation of code changes.
However, we identified two limitations of CC2Vec that need to be addressed: 1) it considers only coarse-grained information about code changes, and 2) it relies on log messages to supervise the training, rather than the self-contained content of the code changes.
We further discuss these issues in detail in Section~\ref{sec:cc2vec}.

To mitigate the above limitations, we present CCBERT, a Transformer-based pre-trained model for obtaining representations for code changes.
CCBERT enriches the input by considering the edit action of each token which is responsible for modifications made to a source code file.
We consider four standard types of edit actions, which are \textit{insert}, \textit{delete}, \textit{replace}, and \textit{equal}.
The changed and unchanged tokens in the changed code lines can be distinguished by considering the edit action.
Besides, to enable training CCBERT on the self-contained content of the code changes, we designed four novel self-supervised pre-training tasks which are \textit{Masked Token Unit Prediction}, \textit{Masked New Token Prediction}, \textit{Masked Old Token Prediction}, and \textit{Masked Edit Action Prediction}. These pre-training tasks are specialized for learning code change representations in a self-supervised way, at the same time bringing the fine-grained edit actions into account.

We evaluate the effectiveness of CCBERT on three downstream tasks, which are just-in-time defect prediction~\cite{DeepJIT}, patch correctness prediction~\cite{tian2020evaluating}, and bug-fixing commit prediction~\cite{patchnet}.
Results show that fine-tuned CCBERT achieves state-of-the-art performance on all three tasks. Specifically, CCBERT significantly outperforms CC2Vec or the state-of-the-art approaches of the downstream tasks by 7.7\%--14.0\% in terms of different metrics.
To further explore the advantages of CCBERT compared to larger pre-trained code models, we thoroughly compared CCBERT with other larger pre-trained code models (CodeBERT~\cite{CodeBERT} and GraphCodeBERT~\cite{GraphCodeBERT}) in terms of training time, inference time, model size, GPU memory needed, and performance gain. 
Our experimental results showed that CBERT has consistently outperformed the other larger pre-trained code models with 6--10 times less training time, 5--30 times less inference time, 3.3 times smaller model size, and 7.9 times less GPU memory needed.

The main contributions of this work are as follows:

\begin{itemize}[leftmargin=*]
    \item We present a Transformer-based self-supervised representation model CCBERT for code change representation that perceives fine-grained code changes at the token level. 
    \item We propose a novel initial representation for each token from raw code changes that considers not only the corresponding pair of old and new versions of the token but also the edit action. Besides, we propose a set of novel self-supervised tasks to pre-train CCBERT.
    \item In our evaluation, CCBERT consistently outperforms CC2Vec and other pre-trained code representation models for three code change-based tasks.
\end{itemize}

\section{Background}
\label{sec:background}

\subsection{Model for Code Change Representation Learning}\label{sec:cc2vec}
CC2Vec~\cite{cc2vec} is a hierarchical attention network-based code change representation model guided by log messages (i.e., commit messages).
CC2Vec comprehends the structural information of a code change and identifies important aspects of the code change with respect to the log message of the code change. Although CC2Vec supports code change representation, we identified the following limitations of CC2Vec need to be addressed:

\begin{figure}[t] 
\centering 
\includegraphics[width=\linewidth]{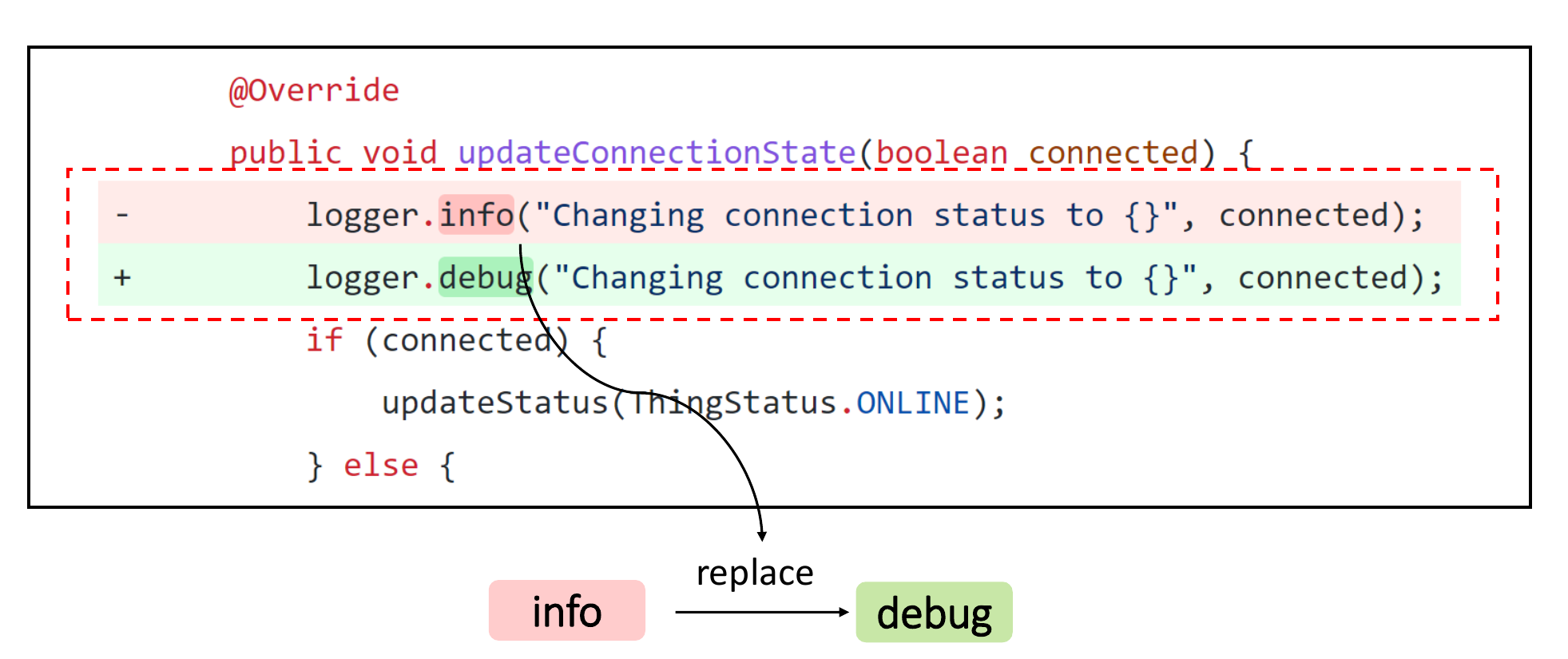} 
\caption{An example
of the difference between the input of CC2Vec and CCBERT. The dashed red (solid black) box is the range of code change data that CC2Vec (CCBERT) considers. CCBERT also emphasizes on the edit action (i.e. \texttt{replace}) and the changed tokens (i.e. \texttt{info} and \texttt{debug}) in the representation learning which are not considered by CC2Vec. 
}
\label{fig:change_example} 
\end{figure}

\begin{itemize}[leftmargin=*]

    \item \textbf{Coarse-grained information.} For a given code change, CC2Vec extracts the changed lines (code in the dashed red box in Figure~\ref{fig:change_example}) from the code changes. Then it sequentially parses the changed code lines into a sequence of tokens without considering fine-grained edit information. 
    Consider the code change shown in Figure~\ref{fig:change_example} as an example: there is only one token changed between the removed and the added code lines, i.e., replace the method \texttt{logger.info} with \texttt{logger.debug}. 
    And the corresponding edit action of the only changed token is \texttt{replace}. In CCBERT, we address this issue by considering the \textit{fine-grained edit actions} on code tokens as input. With the information on edit actions, the intention behind the code change becomes more evident.
    
    \item \textbf{Not trained on the code changes but on log messages.} CC2Vec relies on log messages to supervise its training process. However, many works have demonstrated that log messages are usually poorly written due to the lack of experience and direct motivation, time pressure, and the complexity of code changes~\cite{wang2021quality,liu2018neural,liu2019generating}. Developers often neglect to even write commit messages at all~\cite{dyer2013boa,maalej2010can}. This may impact the quality of the representation learned by CC2Vec. In CCBERT, we mitigate the heavy dependency on log messages by trying to fully leverage the self-contained content of code changes as supervision signals.
    
\end{itemize}

\subsection{Model for Code Representation Learning}
CodeBERT~\cite{CodeBERT} is a bimodal pre-trained model for both programming language (PL) and natural language (NL).
CodeBERT is pre-trained on a large-scale dataset, CodeSearchNet~\cite{codexglue}.
CodeBERT considers two pre-training objectives proposed in natural language processing: masked language modeling (MLM) and replaced token detection (RTD). MLM randomly masks 15\% of tokens in the input and the goal is to predict the masked tokens. 
RTD randomly replaces tokens in the input and trains a model to identify the replaced tokens.
GraphCodeBERT~\cite{GraphCodeBERT} is an extension of CodeBERT that additionally brings the inherent structure of code into consideration. 
Apart from utilizing the traditional MLM task at the pre-training stage, GraphCodeBERT introduces two structure-aware pre-training tasks: edge prediction and node alignment. 
Our work is different from those pre-trained models for code snippets since CCBERT is specifically designed for understanding code changes.

\section{The Proposed Approach}
\label{sec:model}

Figure~\ref{fig:framework} illustrates the overall framework of CCBERT that takes code changes as input and outputs their corresponding distributed representations.
More specifically, CCBERT consists of four parts:

\begin{itemize}[leftmargin=*]
    \item \textbf{Pre-processing.} (Section~\ref{subsec:preprocess}). This part extracts three pieces of fine-grained information (i.e., the old version of code tokens, the new version of code tokens, and the edit actions) from each of the code change and represents each piece as a sequence of tokens. In other words, CCBERT considers not only code change itself but also the corresponding context (i.e., the surrounding unchanged code lines) with fine-grained information (i.e., edit actions) as its input.
    \item \textbf{Input Layer.} (Section~\ref{subsec:il}). This part first encodes the three pre-processed sequences of tokens as well as positional information into a distributed representation format, 2D matrices.
    And then it aggregates the embeddings of four pieces of fine-grained information to form the input to the neural networks in the feature extraction layers.
    \item \textbf{Feature Extraction Layers.} (Section~\ref{subsec:fel}). This part extracts the feature vector of each token unit which corresponds to the old and new versions of the code tokens, edit action, and position by a standard Transformer-based model.
    \item \textbf{Pre-training Objectives.} (Section~\ref{sec:pre-trainOBJ}). 
    This part predicts the labels generated by four proposed pre-training objectives. It enables CCBERT to learn the representation of code changes in a self-supervised manner. \textit{The novel objectives are carefully designed to learn the fine-grained information of code changes from various perspectives.}
    
\end{itemize}

\begin{figure}[t] 
\centering 
\includegraphics[width=\linewidth]{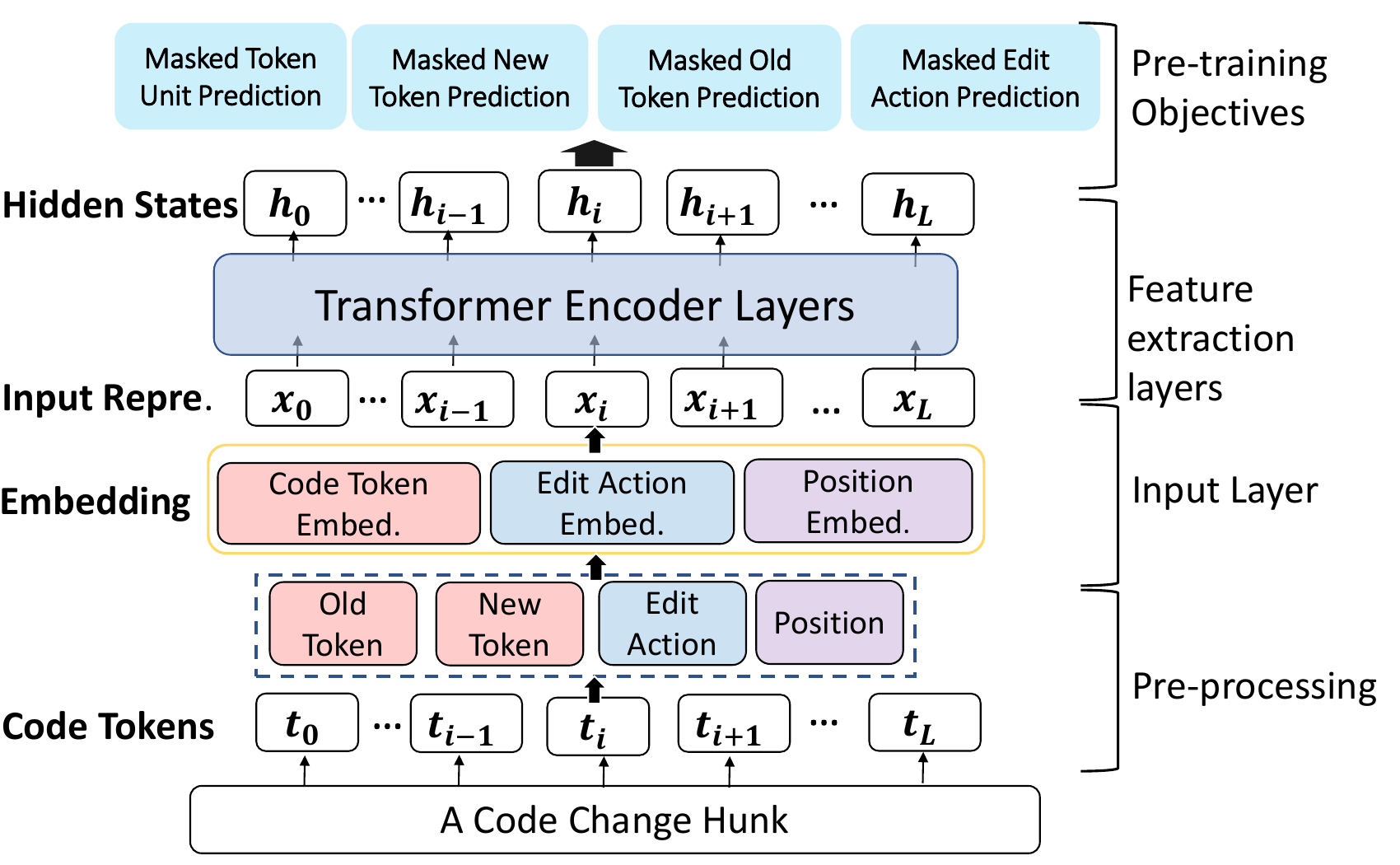} 
\caption{Overall Framework of CCBERT}
\label{fig:framework} 
\end{figure}

\subsection{Pre-processing}\label{subsec:preprocess}

To enable CCBERT to perceive fine-grained information as well as the context of code changes, the first step is to pre-process the corresponding raw data.
Hence, in the pre-processing step, we collect both code changes and their corresponding context (i.e., the surrounding unchanged code lines) and then extract the fine-grained information from them.
In the rest of this section, we introduce the details of how to perform pre-processing gradually as well as extract the edit actions from given code changes particularly.

A commit includes changes made to at least one or multiple files. 
Each changed file contains sets of removed and/or added lines in the commit.
The changes in a file can be grouped into {\em change hunks}. 
In addition to changed lines, a hunk also contains the surrounding unchanged lines to provide context for changes. 
In CCBERT, we follow the default setting of the widely used “git-diff”~\footnote{\url{https://git-scm.com/docs/git-diff}} command to choose \textit{three lines} of unchanged code before and after the changed lines as \textit{context}.
If the distance between any two changed parts is farther than 3 lines, these two changed parts will be split into two hunks.

A commit containing multiple code change hunks can serve tangled purposes~\cite{herzig2013impact} while a single change hunk contains a clearer and simpler intention that is easier for the model to capture.
Therefore, we consider each code change hunk as a data instance fed into CCBERT.
In other words, CCBERT processes code changes at the hunk level.

We process the change hunks by the following steps:

\begin{enumerate}[leftmargin=*]
\item Split code change hunk into two versions of code snippets: the version before the code change (i.e., the old version) and the version after the code change has been applied (i.e., the new version). 
\item Tokenize the two versions of code snippets into two code token sequences. We leverage the well-known Byte-Pair Encoding (BPE) tokenizer~\cite{sennrich2015neural}, which builds the vocabulary by iteratively adding the most frequent combinations of characters and outputs a sequence of token sequences.
For a fair comparison, we reuse the same vocabulary and tokenizer~\footnote{\url{https://huggingface.co/microsoft/codebert-base}} of CodeBERT~\cite{CodeBERT} in CCBERT.
\item To better capture the \textbf{fine-grained changes} between two versions of code snippets, we align two code token sequences and generate the edit sequence by using difflib~\footnote{\url{https://docs.python.org/3/library/difflib.html}}, a module in Python standard library. Figure~\ref{fig:edit_exampl} shows an example to generate an edit sequence.
\end{enumerate}

To generate the edit sequences, difflib finds the longest contiguous matching sub-sequence and computes the edit sequence based on the difference between the longest matching subsequence and each of the two sequences.
We define four different edit actions: \texttt{equal}, \texttt{delete}, \texttt{insert}, and \texttt{replace}.
The \texttt{equal} action indicates tokens in two versions at the same position are the same. The \texttt{insert} and \texttt{delete} actions mean that a new token is added and that an existing token is removed, respectively. 
The \texttt{replace} action denotes that an old token is replaced with a new one.
Specifically, for \texttt{insert} and \texttt{delete} actions, we define a special token ``\texttt{<NULL>}'' to represent an empty token position. For a position that needed an \texttt{insert} action, the old version token is ``\texttt{<NULL>}'' and the new version token is the token inserted. Reversely, for the \texttt{delete} action, the old version token is the token that is deleted, and the new version token is ``\texttt{<NULL>}''. 
Figure~\ref{fig:edit_exampl} illustrates an example of the generated edit sequence, indicating that the exact fine-grained changes are: replacing the code token ``\texttt{remove}'' with ``\texttt{add}'' and inserting ``\texttt{New}'' before ``\texttt{Things}''. The edit sequence can highlight the fine-grained code changes at the token level for CCBERT to learn.

At the end of the pre-processing steps, a code change hunk is converted into three sequences corresponding to the old version code tokens, the new version code tokens, and the edit actions. 
To clarify, we define an \emph{input unit} (also called \emph{token unit}) by combining the old and new versions of the code token as well as the corresponding edit action at the same position.

\begin{figure}[t] 
\centering 
\includegraphics[width=\linewidth]{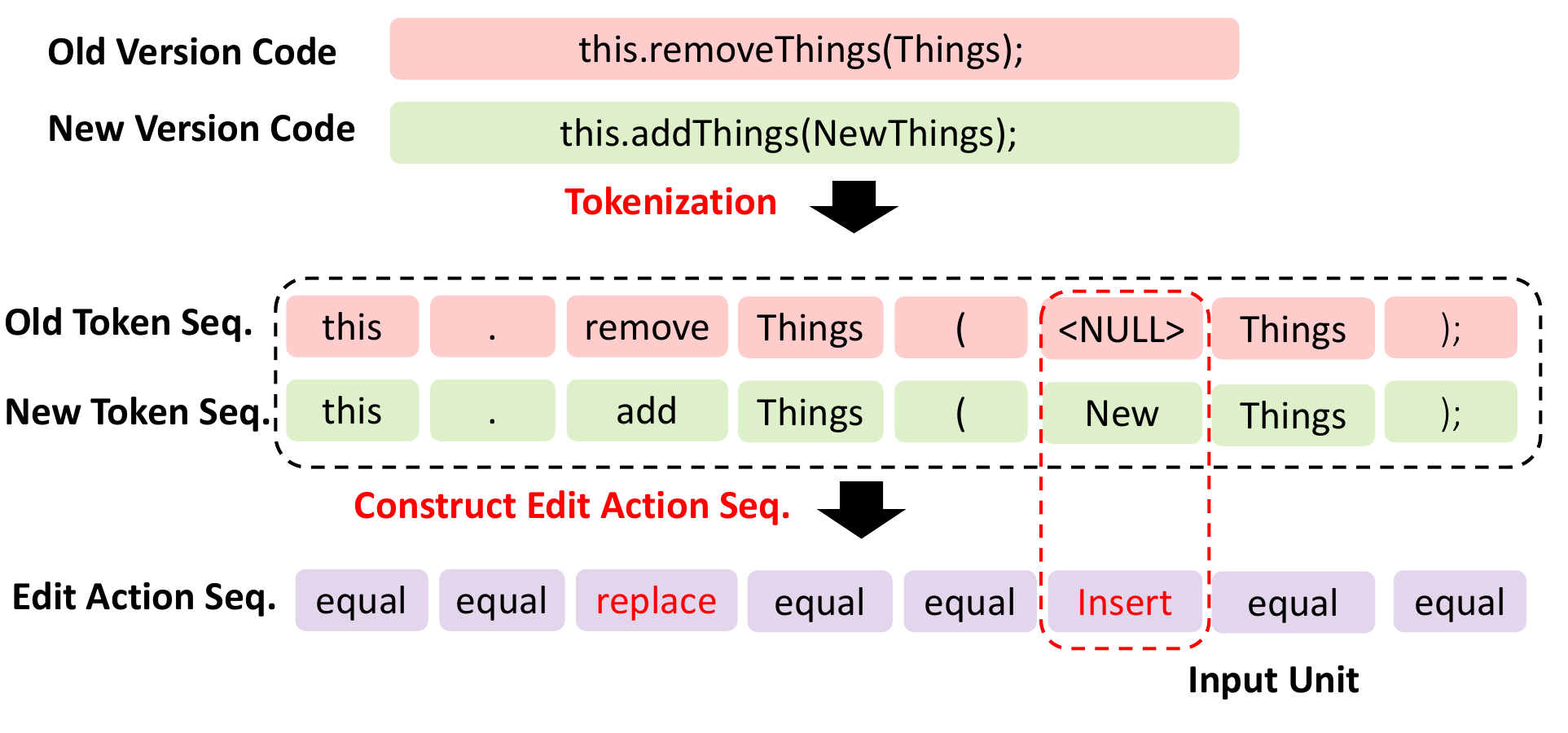} \vspace{-0.5cm}
\caption{Example of generating the edit action sequence.
}
\label{fig:edit_exampl} 
\end{figure}

\subsection{Input Layer}\label{subsec:il}

CCBERT utilizes the edit action sequences to capture the fine-grained code changes. Thus, in the Input Layer, we need to vectorize not only the code tokens but also the edit actions. To better aggregate diverse information extracted from code changes in the pre-processing, we build input representations that fuse the embeddings of old\&new versions of code tokens, edit actions, and the positions of tokens.
The input layer mainly performs the following steps:

\noindent{\textbf{Vectorize Tokens (Edits).}}
We represent each code token or edit action in the pre-processed sequences as a vector of $d$ dimensions. By default, we set $d$ to be 512. 
We randomly initialize two matrices for code tokens and edit actions: the code token matrix $C \in R^{V_{code} \times d}$ and the edit action matrix $E \in R^{V_{edit} \times d}$, where $V_{code}$ is the number of unique code tokens in the vocabulary and $V_{edit}$ is the number of unique edit actions.
Specifically, $C^{i} \in R^{d}$ is the embedding of $i$-th code token in the vocabulary.
The code token matrix and the edit action matrix will be updated during the training process via the back-propagation mechanism~\cite{hagan1994training}.

Given three pre-processed sequences of length $L$, we encode the sequences into three matrices representations by retrieving the corresponding embeddings from $C$ and $E$.
We denote the old code, new code, and the edit actions in a code change hunk as
$X_{old}^{L} \in R^{L \times d}$, $X_{new}^{L} \in R^{L \times d}$, and $ X_{edit}^{L} \in R^{L \times d}$, where $d$ is the dimension of the token (edit) representation.

\noindent{\textbf{Vectorize Position Information.}} 
As CCBERT is built upon BERT~\cite{bert},
it follows the default BERT position embeddings to capture the positional information in the sequence.
We denote the positional embedding as $X_{P}^{L} \in  R^{L \times d}$ when the length of a sequence is $L$.

\noindent{\textbf{Build Input representation.}} After getting the embeddings of code tokens (new or old), edit actions, and positions, the input representation is computed by element-wise summing the embeddings of corresponding old tokens, new tokens, edit actions, and positions: $X^{L} = X_{old}^{L} + X_{new}^{L} + X_{edit}^{L} + X_{P}^{L}$. Hereafter, we refer to $X^{L}$ as $X$ for simplicity.
Besides, a token (input) unit is defined as the combination of the old, new version code token, and the corresponding edit action at the same position. Thus, the representation of the $i$-th token unit is $X^{i} \in R^{d}$ which is the element-wise sum of the old and new code token embedding, the edit action embedding, and the positional embedding at the $i$-th position.
CCBERT employs this element-wise addition of vectors to achieve a shorter input length and reduce computation costs. CCBERT leverages a Transformer encoder, where the maximum input length is a crucial factor affecting the model size and GPU resources required for training. The attention matrix computations in the Transformer scale quadratically with the input sequence length~\cite{10.1145/3530811}.
Element-wise addition has the advantage of keeping the input lengths shorter. For example, if each of the three vectors $X_{old}^{L}$, $X_{new}^{L}$, and $X_{edit}^{L}$ consists of 400 tokens, the final input length remains at 400 tokens using the element-wise addition. On the other hand, if the edit tokens were concatenated separately to the code token sequence, the input length would double to 800 tokens, resulting in significantly higher computation costs.

\subsection{Feature Extraction Layers}\label{subsec:fel}

CCBERT utilizes the Transformer encoder~\cite{transformer} model to extract the features from code changes.
The Transformer encoder~\cite{transformer} with its self-attention mechanism can update the representation of each token unit (i.e., $X^{i}$) by aggregating information from other token units (containing information of code tokens and edit actions at other positions). In this way, the representation of each token unit carries not only its own information but also the overall information of the whole code changes.

A Transformer encoder model usually consists of $N$ identical Transformer encoder layers.
The input of the next transformer encoder layer is the output of the last transformer encoder layer. We use the equations below to illustrate it: $X_{k+1} = TransformerEncoderLayer_k(X_{k})$ where $X_{k}$ is the input matrix for the $k$-th transformer encoder layer. The input vector $X$ (the code change hunk vectorized by the input layer) is fed into the first transformer encoder layer ($X = X_{1}$).

The number of transformer encoder layers (i.e., $N$) significantly affects the model size, training time, and computing resources required for pre-trained models~\cite{bert}. Although more layers of transformer encoder layers usually lead to better performance,
we choose the $N$ as 4 considering the efficiency of our representation model and the limited computation resources (i.e., GPU memory) available in our group.

\subsection{Pre-training Objectives}\label{sec:pre-trainOBJ}

\begin{figure}[t] 
\centering 
\includegraphics[width=0.35\linewidth]{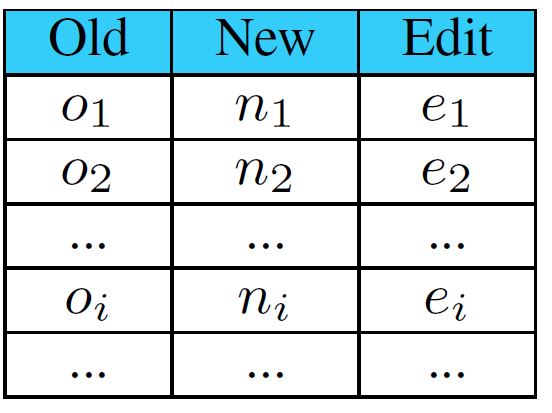} 
\caption{An abstracted example of code change data
}
\label{table:example} 
\end{figure}

\noindent{\textbf{Motivation.}}
Although there is a large scale of code change available in open-source software, it is usually unlabeled which means the properties of a code change related to downstream tasks are unknown. For instance, we may not be able to be aware of whether a code change introduces a defect or not. To learn the semantic meanings of unlabeled data, \emph{a typical idea is first to mask some parts of unlabeled data and then ask the DL models to reconstruct the masked parts}. If a model can accurately reconstruct the masked parts of input data, it indicates that the model captures some core features of this kind of input data~\cite{bert}. 
To learn an effective code change representation model on a large unlabelled code change data, one core question is: \emph{what parts of code change data can we mask?}
As introduced in Section~\ref{subsec:preprocess}, code change data can be pre-processed into three sequences of tokens: the old version code tokens, the new version code tokens, and the edit actions. Our pre-training objectives aim to fully use the three kinds of information.

Figure~\ref{table:example} presents an abstracted example. It presents the old version code tokens (the first column), the new version code tokens (the second column), and the edit actions (the third column). 
Specifically, in Figure~\ref{table:example}, $o_i$, $n_i$, and $e_i$ stand for the token at the $i$-th positions in three sequences.
To fully make use of all components to help CCBERT to perceive the semantics of code changes, we can hide/mask the following complementary parts of code changes:
\begin{itemize}[leftmargin=*]
    \item Hide All at Certain Positions: randomly mask all tokens (two versions of code tokens and edit actions) at certain positions (e.g., masking $<o_i, n_i, e_i>$);
    \item Hide New: mask the version of code after a code change (i.e., masking a number of the $n_i$);
    \item Hide Old: mask the version of code before a code change (i.e., masking a number of the $o_i$);
    \item Hide Edit: mask the edit actions in a code change (e.g., masking a number of the $e_i$).
\end{itemize}
An overall example of four tasks is presented in Figure~\ref{fig:objectives}. We will introduce more details about our pre-training objectives in the following parts.

\begin{figure}[t] 
\centering 
\includegraphics[width=\linewidth]{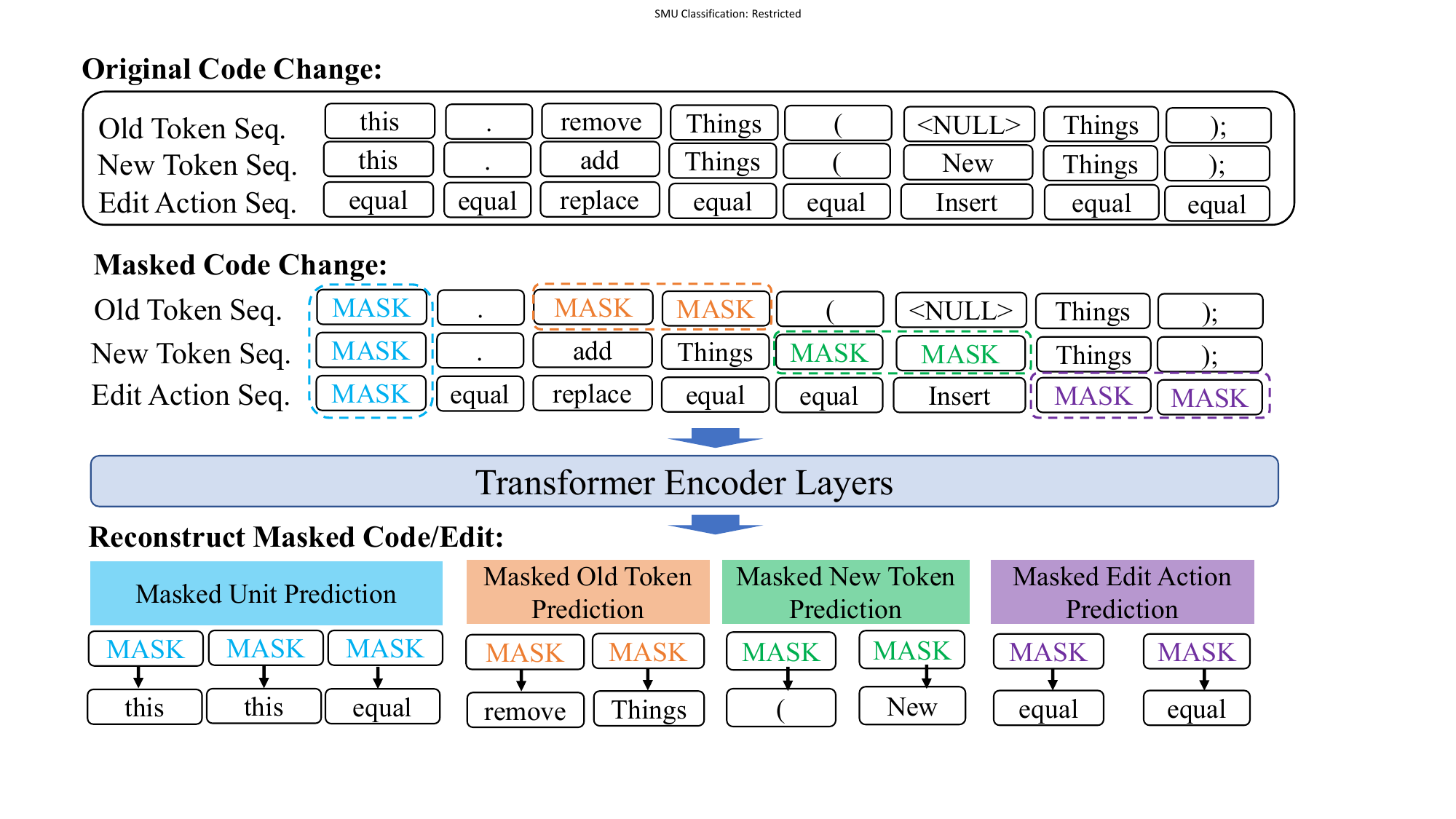}
\caption{Pre-training Objectives Examples. The dashed blue box indicates the masking of Masked Unit Prediction (MUP) and the Transformer encoder will reconstruct the parts that are masked in MUP. Similarly, the dashed orange, green, and purple boxes stand for Masked Old Token Prediction, Masked New Token Prediction, and Masked Edit Action Prediction Respectively.}
\label{fig:objectives} 
\end{figure}

\vspace{0.2cm}
\noindent{\textbf{Masked Token Unit Prediction.}}
This pre-training objective is based on the ``Hide All at Certain Positions'' strategy. It masks the token units $\langle o_i, n_i, e_i \rangle$ that carry three pieces of information, old code token, new code token, and edit action, at the same time. Then it asks CCBERT to predict the masked token units.
As the code sequences (old and new) and the edit sequences are aligned (as shown in Figure~\ref{fig:edit_exampl}), we only need to decide the positions we want to mask. Following the previous work~\cite{bert,CodeBERT}, we randomly sample a standard portion (i.e., 15\%) of positions to mask.
Once the masked positions are determined, we replace the code tokens and the edit actions at the masked positions with a special token ``\texttt{<MASK>}''.
After masking a part of input data, CCBERT is required to predict what code tokens/edit actions are masked in inputs.
We describe the loss of the \underline{M}asked Token \underline{U}nit \underline{P}rediction (MUP) task as:
\begin{equation*}
\begin{aligned}
    L_{MUP} = - \sum_{t=1}^{K} log \{p_{\theta } (n_{q_{1}(t)}|n^{masked}, o^{masked}, e^{masked}) \\
    \times p_{\theta } (o_{q_{1}(t)}|n^{masked}, o^{masked}, e^{masked}) \\
    \times p_{\theta } (e_{q_{1}(t)}|n^{masked}, o^{masked}, e^{masked})\}
\end{aligned}
\end{equation*}

where $K$ is the number of masked token units in this data sample. $n^{masked}$, $o^{masked}$, $e^{masked}$ are the token/edit sequences of the masked new version of code, the masked old version of code, and the masked edit actions respectively.  
$q_{1}(t)$ is the real position of the $t$-th masked new token, old token, and edit action in its corresponding sequences. $q_{1}$ is a list whose elements are randomly sampled 15\% positions of input sequences.

\vspace{0.2cm}
\noindent{\textbf{Masked New/Old Token Prediction.}}
These two pre-training objectives are based on the ``Hide New'' and ``Hide Old'' strategies.
We expect the model to imitate what human developers have done when developing software (i.e. editing the old version of code into a new version), which motivates us to propose the objective named Masked New Token Prediction. The objective requires a model to predict the new version code based on the given old version code.
It is challenging to achieve as the purposes of code changes could vary greatly.
Such properties introduce randomness that may hinder the pre-training.
To mitigate this issue, we not only feed the old version of code tokens but also the corresponding edit actions.
Specifically, we randomly sample a portion (i.e., 15\%) of the new version of code tokens to mask.

To fully use all the information in a code change, we also randomly mask 15\% of the old version of tokens and retain the new version of tokens unmasked. We call this pre-training objective Masked Old Token Prediction which requires the DL model to ``roll back'' the old version of code from the new version of code and corresponding edit actions.
We describe the loss of the \underline{M}asked \underline{N}ew Token \underline{P}rediction (MNP) task and the \underline{M}asked \underline{O}ld Token \underline{P}rediction (MOP) task as:
\begin{equation*}
    L_{MNP} = - \sum_{t=1}^{K} log p_{\theta } (n_{q_{2}(t)}|n^{masked}, o, e)
\end{equation*}

\begin{equation*}
    L_{MOP} = - \sum_{t=1}^{K} log p_{\theta } (o_{q_{3}(t)}|n, o^{masked}, e)
\end{equation*}

where $K$ is the number of masked new(old) code tokens. $n^{masked}$($o^{masked}$) is the masked new(old) code token sequence and $n_i$ ($o_i$) is the masked new(old) version of code token at the $i$-th position. $q_{2}(t)$ and $q_{3}(t)$ are the real positions of the $t$-th masked new and old tokens in their corresponding code token sequences respectively.  $q_{2}$ and $q_{3}$ are two lists whose elements are randomly sampled 15\% positions of input sequences. Please note that $q_{1}$, $q_{2}$, and $q_{3}$ are all randomly sampled with different random seeds.

\vspace{0.2cm}
\noindent{\textbf{Masked Edit Action Prediction.}}
This pre-training objective is based on the ``Hide Edit'' strategy. 
It is also critical for the model to understand the real meaning of edit actions like ``insert'', ``delete'', or ``replace''.
To enable CCBERT to perceive the meaning of edit actions, we only mask edit actions in a code change hunk without masking any new or old code tokens. 
The objective is to predict the type of each masked edit action by giving two versions of the code tokens.  
As this pre-training objective is relatively simpler than the other three objectives, we do not randomly mask 15\% of edit actions. Instead, we mask \textit{all} the edit actions in a code change hunk and require the model to predict each edit action on each pair of old and new code tokens.
We describe the loss of the \underline{M}asked \underline{E}dit Action \underline{P}rediction (MEP) task as:
\begin{equation*}
    L_{MEP} = -\sum_{t=1}^{|e^{masked}|} log p_{\theta } (e_{t}|n, o, e^{masked})
\end{equation*}
where $e^{masked}$ is the masked edit action sequence and $|e^{masked}|$ is the length of the masked edit action sequence. $e_t$ is the masked edit action at the $t$-th position.

\subsection{Parameter Learning}
During the pre-training stage, CCBERT learns the following parameters: the embedding matrices of code tokens (i.e., $C$) and edit actions (i.e., $E$), the transformer encoder layers’ matrices, 
and the weights matrices and bias values of the fully connected layers. 
Once the parameters are learned, the representation of each code change hunk is determined.
The goal of the pre-training stage is to minimize the sum of the losses of four pre-training objectives and the model parameters are shared among all the objectives. 
The final loss function is given below:
\begin{equation*}
\min_{\theta} \mathcal{L}_{MUP} (\theta) +  \mathcal{L}_{MNP}(\theta) + \mathcal{L}_{MOP}(\theta) + \mathcal{L}_{MEP}(\theta)
\end{equation*}
where $\theta$ represents all parameters of CCBERT.
$\mathcal{L}_{MUP}$, $\mathcal{L}_{MNP}$, $\mathcal{L}_{MOP}$, and $\mathcal{L}_{MEP}$ refer to the losses of Masked Token Unit Prediction, Masked New Token Prediction, Masked Old Token Prediction, and Masked Edit Action Prediction objectives.

\section{Experiments}
\label{sec:experiment}

\subsection{Task-Agnostic Experimental Details}

\subsubsection{Pre-training Data}
We pre-train CCBERT on a large-scale code change dataset recently collected by Monperrus et al.~\cite{megadiff}, which includes 663k commits in Java projects on Github.
This dataset is built based on 101,472 unique open-source GitHub repositories and filtered with a set of constraints.
 For example, the dataset only included
(1) commits that contained changes in at least one Java source file, and 
(2) commits that contained less than $n$ (by default $n=40$) changed lines of code in a Java source file.
Using the preprocessing process described in Section~\ref{subsec:preprocess}, we split these commits into over 1.3M change hunks.
Although we only pre-trained on Java data which is one of the most popular programming languages, our approach is generic and language-agnostic and can be applied to any programming language.

\subsubsection{Pre-training Implementation}
We implemented our model using PyTorch~\cite{paszke2017automatic}.
In addition, we adopted the Transformer architecture from the UER library~\cite{zhao2019uer}. 
We trained CCBERT on NVIDIA Tesla V100 GPU with 16 GB of memory. 
The hyper-parameters to train models are as follows: the batch size is 64 and the learning rate is $3\mathrm{e}{-4}$.
AdamW~\cite{loshchilov2018decoupled} was used to update the parameters and we set the number of warm-up steps as 10K. We set the max input length as 256 due to the limitation of the computation resources\footnote{To avoid the Out Of Memory (OOM) issue, we had to reduce the maximum input length from 512 to a smaller value that was compatible with the device's capacity.}. We set the max training step as 500K.

\subsubsection{Fine-tuning Implementation}
\label{sec:finetune}
For each task, we simply feed the task-specific inputs and outputs into CCBERT and fine-tune all the parameters of CCBERT. Specifically, 
the average representation in the last layer is fed into a linear classifier to produce the prediction. A standard cross entropy
loss is leveraged and the parameters of CCBERT are updated to minimize the cross entropy loss function.
We evaluate the performance of the model at the end of each epoch on the validation set and select the best-performing model for testing.

\subsection{Research Question}

The goal of this work is to build a pre-trained representation model of code changes that can be applied to multiple downstream tasks. 
We raise a main research question: 
\textbf{\textit{How does CCBERT perform in downstream tasks?}}

We evaluate the effectiveness of CCBERT on three different code change-based tasks, i.e., just-in-time defect prediction~\cite{DeepJIT}, patch correctness prediction~\cite{tian2020evaluating}, and bug-fixing commit prediction~\cite{patchnet}.

\subsection{Task 1: Just-in-time Defect Prediction}

\subsubsection{Background}
The task of just-in-time (JIT) defect prediction refers to the identification of defective code change, which can provide instant feedback to developers to minimize their effort for inspection. JIT defect prediction tools have been widely spread in large software companies~\cite{mockus2000predicting, shihab2012industrial}. We model the task as a binary classification task, labeling a commit as defective or not.

\subsubsection{Baselines}
We adopt CC2Vec~\cite{cc2vec}, the state-of-the-art code change representation model, as one baseline. 
Moreover, we adopt two recent task-specific approaches designed for the JIT defect prediction task: LApredict~\cite{Zeng2021DeepJD} and DeepJIT~\cite{DeepJIT} as baselines.
LApredict leverages the number of added lines in a commit as the input and builds a logistic regression model as the classifier.
DeepJIT~\cite{DeepJIT} is a CNN-based model~\cite{deeplearning} that automatically extracts features from the content of code changes and uses FCs as the classifier.

\subsubsection{Experimental Setting}
As CCBERT was only pre-trained on Java code changes provided in the dataset~\cite{megadiff}, 
we selected Java as our target language for both training and evaluation.
For the datasets, we use the Java JIT defect prediction datasets used in the LApredict paper~\cite{LApredit}.  
We follow the same data partitioning strategy used in the LAPredict
work~\cite{LApredit}. The first 80\% of data in chronological order are the training set and the later 20\% of data are the testing set. We use the same data split for training and evaluating all the models investigated.  
We set aside 5\% of data from training data as the validation set. Therefore, the data split ratio is 75\%, 5\%, and 20\% for training, validation, and testing, respectively.  We keep the hyper-parameters the same as the original settings reported in the corresponding papers for each baseline.

The datasets released in LApredict are imbalanced in classes: the number of defective commits is smaller than clean commits. Specifically, the ratios between the number of defective commits and clean commits are about 1:10,  2:3, and 4:6 in Gerrit, Platform, and JDT respectively. We follow prior work~\cite{DeepJIT, cc2vec, LApredit} to avoid using threshold-dependent measures (e.g., accuracy or F1-scores) since these measures strongly depend on arbitrary thresholds~\cite{nguyen2009learning, gu2008data}. 
We use Area Under Receiver Operator Characteristic Curves (AUC-ROC) as the evaluation metric, which is a threshold-independent measure and is commonly used in previous studies on JIT defect prediction~\cite{DeepJIT, cc2vec, LApredit}. AUC-ROC shows how the number of correctly classified positive examples varies with the number of misclassified negative examples.
However, ROC curves can present an overly optimistic view of the performance of a model if there is a skew in the class distribution~\cite{davis2006relationship}. To mitigate the problem, we also adopt Area Under Receiver Precision-Recall Curves (AUC-PR) which is more informative than AUC-ROC when evaluating binary classifiers on imbalanced datasets~\cite{saito2015precision, davis2006relation}.

\begin{table}[t]
\caption{Comparison in JIT Defect Prediction}
\label{table:task1}
\resizebox{\columnwidth}{!}{%
\begin{tabular}{l|l|r|r|r|r}
\hline
\textbf{Datasets}                  & \textbf{Metric} & \multicolumn{1}{c|}{\textbf{CC2Vec}} & \multicolumn{1}{c|}{\textbf{LApred.}} & \multicolumn{1}{c|}{\textbf{DeepJIT}} & \multicolumn{1}{c}{\textbf{CCBERT}} \\ \hline
\multirow{2}{*}{\textbf{JDT}}      & PR           & 64.0                                & 64.4                                   & 64.4                                 & \textbf{67.4}                      \\ \cline{3-6} 
                                   & ROC          & 66.5                                & 67.7                                   & 67.0                                 & \textbf{70.6}                      \\ \hline
\multirow{2}{*}{\textbf{Platform}} & PR           & 55.9                                & 53.3                                   & 57.4                                 & \textbf{67.9}                      \\ \cline{2-6} 
                                   & ROC          & 76.1                                & 74.7                                   & 77.1                                 & \textbf{83.0}                      \\ \hline
\multirow{2}{*}{\textbf{Gerrit}}   & PR           & 13.4                                & \textbf{18.4}                          & 13.5                                 & 16.4                               \\ \cline{2-6} 
                                   & ROC          & 69.9                                & 75.0                                   & 70.3                                 & \textbf{76.0}                      \\ \hline
\multirow{2}{*}{\textbf{Average}}  & PR           & 44.4                                & 45.3                                   & 45.1                                 & \textbf{50.6}                      \\ \cline{2-6} 
                                   & ROC          & 70.8                                & 72.5                                   & 71.5                                 & \textbf{76.5}                      \\ \hline
\end{tabular}
}
\end{table}

\subsubsection{Results}
Table~\ref{table:task1} presents the experimental results for the JIT defect prediction task.
CCBERT achieves the best performance for all projects except for the AUC-PR on the Gerrit project. 
It outperforms DeepJIT (LApredict)  by 7.0\% and 12.2\% (5.5\% and 11.7\%) in terms of AUC-ROC and AUC-PR respectively on average.
In addition, CCBERT outperforms CC2Vec by 8.1\% and 14.0\% in terms of AUC-ROC and AUC-PR.
To ensure the improvements are statistically significant, we conducted the Wilcoxon Signed Rank Test~\cite{wilcoxon1992individual} at a 95\% confidence level (i.e., $p \text{-} value < 0.05$) on the paired data which corresponds to each of the baselines and CCBERT.
We confirmed that the difference between all baselines and CCBERT is statistically significant.

\subsection{Task 2: Patch Correctness Prediction}

\subsubsection{Background}
The task of patch correctness prediction aims to identify the correct patches generated by automated program repair (APR) approaches~\cite{tian2020evaluating}. 
A typical way to assess the correctness of a generated patch is to validate it by executing available test cases. 
However, this patch validation method suffers from an overfitting problem~\cite{qi2015analysis,smith2015cure}:
the generated patches, although validated by passing all available test cases, may actually be incorrect with respect to the intended program specification. 
To mitigate the overfitting problem, 
recently, Tian et al.~\cite{tian2020evaluating} showed the potential of code representations to predict the correctness of patches. Later, Building upon this idea, Lin et al.\cite{CACHE} extended this research direction further. 
We follow Tian et al.~\cite{tian2020evaluating} and Lin et al.~\cite{CACHE} to define the patch correctness prediction task as a binary classification task: given buggy and patched code, a model should predict whether the patched code can correctly fix the bug or not.

\subsubsection{Baselines}
We adopt CC2Vec~\cite{cc2vec}, the state-of-the-art code change representation model, as one baseline. 
Moreover, we adopt two recent task-specific approaches designed for this task: 
Tian et al.'s~\cite{tian2020evaluating} approach and CACHE~\cite{CACHE}.
Tian et al. tried different representation models such as Doc2Vec~\cite{doc2vec} and BERT~\cite{bert} with different classifiers like Logistic Regression (LR) and Decision Tree. For limited space, we only report their best-performing combination (i.e., BERT+LR). 
CACHE~\cite{CACHE} is proposed by Lin et al. and showed state-of-the-art performance in the patch correctness assessment task. Specifically, CACHE learns a context-aware code change embedding considering program structures.

\subsubsection{Experimental Setting}
For the datasets, we use the dataset collected by Tian et al.~\cite{tian2020evaluating}.
The dataset contained 1,000 patches and was built based on labeled patches provided by two independent teams: Liu et al.~\cite{liu2020efficiency} and Xiong et al.~\cite{xiong2018identifying}.
We follow the same data partitioning used by Tian et al.~\cite{tian2020evaluating}, i.e., 5-fold cross-validation.
For evaluation metrics, as the dataset released by Tian et al.~\cite{tian2020evaluating} is balanced, we adopt both threshold-independent metrics like AUC-ROC and threshold-dependent metrics in this task.
Similar to the previous studies~\cite{tian2020evaluating, le2021deepcva}, we use the metrics that are suitable to evaluate binary classification from diverse perspectives: accuracy, F1-score, and AUC-ROC.

\begin{table}[]
\caption{Comparison in Patch Correctness Prediction}
\centering
\small
\begin{tabular}{llll}
\hline
\multicolumn{1}{l|}{\textbf{Methods}}      & \multicolumn{1}{c|}{\textbf{Acc.}} & \multicolumn{1}{c|}{\textbf{F1}}   & \multicolumn{1}{c}{\textbf{AUC}}    \\ \hline
\multicolumn{1}{l|}{\textbf{CC2Vec}~\cite{cc2vec}}  & \multicolumn{1}{l|}{73.9}          & \multicolumn{1}{l|}{72.0}          & \multicolumn{1}{l}{78.8}                  \\ \hline
\multicolumn{1}{l|}{\textbf{Tian et al.}~\cite{tian2020evaluating}}    & \multicolumn{1}{l|}{74.4}          & \multicolumn{1}{l|}{72.0}          & \multicolumn{1}{l}{80.8}                 \\ \hline
\multicolumn{1}{l|}{\textbf{CACHE}~\cite{CACHE}}       & \multicolumn{1}{l|}{75.4} & \multicolumn{1}{l|}{78.0} & \multicolumn{1}{l}{80.3}   \\ \hline
\multicolumn{1}{l|}{\textbf{CCBERT}}       & \multicolumn{1}{l|}{\textbf{81.0}} & \multicolumn{1}{l|}{\textbf{80.0}} & \multicolumn{1}{l}{\textbf{88.4}}  \\ \hline
\end{tabular}
\label{task_two}
\end{table}

\subsubsection{Results}
The performance of the different approaches is presented in Table~\ref{task_two}.
CCBERT significantly outperforms CC2Vec by 9.6\%, 11.1\%, and 12.2\% in terms of accuracy, f1-score, and AUC-ROC. 
CCBERT outperforms the Tian et al.' approach (i.e., BERT+LR) by 8.9\%, 11.1\%, and 9.4\% in terms of accuracy, F1-score, and AUC-ROC. 
In addition, CCBERT leads to 7.4\%, 2.6\%, and 10.1\% improvements over CACHE in terms of accuracy, F1-score, and AUC-ROC, respectively.
To ensure the improvements are statistically significant, we carry out the Wilcoxon Signed Rank Test~\cite{wilcoxon1992individual} at a 95\% confidence level (i.e., $p \text{-} value < 0.05$) on the paired data which corresponds to each of the baselines and CCBERT. We find that, in statistics, CCBERT significantly outperforms all baselines in this task.

\subsection{Task 3: Bug-fixing Commit Prediction}
\subsubsection{Background}
The Linux kernel follows a two-tiered release model: a \textit{mainline} version that accepts bug fixes and feature enhancements, is paralleled by a series of \textit{stable} versions, which accept only bug fixes~\cite{lee2003firm}.
Developers of the Linux kernel regularly propagate bug-fixing patches in a \textit{mainline} version to the \textit{stable} versions to ensure the quality of stable versions.
However, the maintainers of \textit{stable} versions may overlook relevant patches in the latest \textit{mainline} version. 
Thus, an automated method to identify bug-fixing patches is needed.
We treat the problem as a binary classification problem, in which each patch is labeled as a bug-fixing patch or not.

\subsubsection{Baselines}
We adopt CC2Vec~\cite{cc2vec}, the state-of-the-art code change representation model, as one baseline. 
Moreover, we adopt two task-specific approaches, i.e., LPU-SVM~\cite{tian2012identify} and PatchNet~\cite{patchnet}.
LPU-SVM is proposed by Tian et al.~\cite{tian2012identify}. It uses hand-crafted features as model input and combines Learning from Positive and Unlabeled examples (LPU)~\cite{li2003learn} with Support Vector Machine (SVM)~\cite{cortes1995support} to build a patch classification model. PatchNet~\cite{patchnet} is he state-of-the-art task-specific approach, which represents the patches as a three-dimensional matrix and employs a 3D-CNN~\cite{3dcnn} to automatically extract features from the matrix.

\subsubsection{Experimental Setting}
For the dataset, we use the dataset of Linux kernel bug-fixing patches published by Hoang et al.~\cite{patchnet}. This dataset consists of 42K bug-fixing
patches as well as 40K non-bug-fixing patches collected from the Linux kernel versions v3.0 to v4.12, released in July 2011 and July 2017 respectively. The dataset only considered patches that have less than 100 lines of changed code by following the Linux kernel stable patch guidelines. 
We used a 5-fold cross-validation for the evaluation as Hoang et al. have done~\cite{patchnet}.
Following the previous study~\cite{patchnet, cc2vec}, we use the same metrics to evaluate the bug-fixing commit prediction task (i.e., accuracy, F1-score, AUC-ROC).

\begin{table}[t]
\caption{Comparison in Bug-fixing Commit Prediction}
\small
\centering
\begin{tabular}{l|c|c|c}
\hline
\textbf{Methods}    & \textbf{Acc.} & \textbf{F1}  & \textbf{AUC}    \\ \hline
\textbf{CC2Vec}~\cite{cc2vec}     & 89.3          & 89.3          & 95.3                   \\ \hline
\textbf{LPU-SVM}~\cite{tian2012identify}    & 73.1          & 73.3          & 73.1                      \\ \hline
\textbf{PatchNet}~\cite{patchnet}   & 85.4          & 85.2          & 93.3                    \\ \hline
\textbf{CCBERT}     & \textbf{91.6} & \textbf{91.8} & \textbf{96.8}  \\ \hline
\end{tabular}
\label{Task_3_result}
\end{table}

\subsubsection{Results}
We report the performance of the different approaches in Table~\ref{Task_3_result}. 
Overall, CCBERT shows the best performance in this task.
Specifically, CCBERT outperforms the task-specific state-of-the-art approach,  PatchNet by 7.3\%, 7.7\%, and 3.8\% in terms of accuracy, F1-score, and AUC-ROC.
CCBERT also consistently outperforms CC2Vec by 2.6\%, 2.8\%, and 1.6\% in terms of accuracy, F1-score, and AUC-ROC.
We conducted the Wilcoxon Signed Rank Test~\cite{wilcoxon1992individual} at a 95\% confidence level (i.e., $p \text{-} value < 0.05$). Results show that the improvements achieved by CCBERT are statistically significant.

\section{Discussions}
\label{sec:discussion}

\subsection{Qualitative Analysis}

\begin{figure}[b]
\centering  
\subfigure[\textbf{Vector space learned by CC2Vec.}]{   
\begin{minipage}{4cm}
\centering   
\includegraphics[scale=0.26]{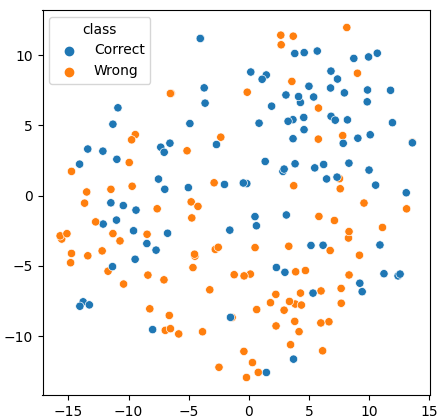}  
\end{minipage}
}
\subfigure[\textbf{Vector space learned by CCBERT.}]{ 
\begin{minipage}{4cm}
\centering    
\includegraphics[scale=0.26]{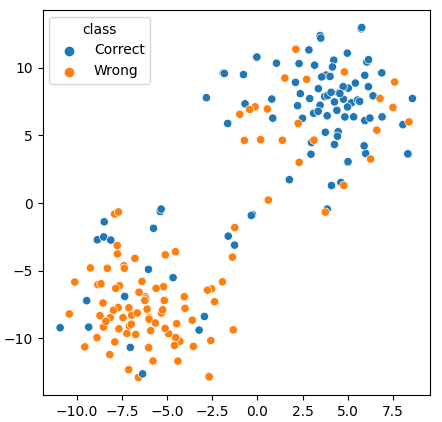}
\end{minipage}
}
\caption{Vector space visualization of CCBERT and CC2Vec in the patch correctness prediction task.}    
\label{fig:visual}   
\end{figure}

To investigate the quality of code change representation learned by CCBERT, we utilize the T-SNE dimensionality reduction technique~\cite{van2008visualizing} to visualize the vector space learned by our approach and CC2Vec in the patch correctness prediction task.
As shown in Figure~\ref{fig:visual},
the code change vectors of CCBERT are more linear-distinguishable than CC2Vec vectors, indicating that CCBERT contains much clearer patch correctness information than CC2Vec. This may explain why CCBERT can significantly outperform the prior state-of-the-art, CC2Vec.

\begin{table}[t]
\centering
\caption{Ablation study on pre-training objectives and comparison with the original MLM. }
\begin{tabular}{l|r|r|r}
\hline
\textbf{Ablation}     & \textbf{\begin{tabular}[c]{@{}r@{}}Task1\\ (ROC)\end{tabular}} & \textbf{\begin{tabular}[c]{@{}r@{}}Task2\\ (F1)\end{tabular}} & \textbf{\begin{tabular}[c]{@{}r@{}}Task3\\ (F1)\end{tabular}} \\ \hline
\textbf{CCBERT}   & \textbf{76.5}                                                  & \textbf{80.0}                                                 & \textbf{91.8}                                                 \\ \hline
- Unit        & 75.1                                                           & 77.9                                                          & 91.8                                                          \\ \hline
- New         & 74.9                                                           & 75.2                                                          & 91.6                                                          \\ \hline
- Old         & 75.7                                                           & 77.4                                                          & 91.8                                                          \\ \hline
- Edit        & 74.7                                                           & 77.0                                                          & 91.6                                                          \\ \hline \hline
\textbf{Original MLM} & 72.9                                                           & 72.3                                                          & 90.7                                                          \\ \hline 
\end{tabular}
\label{table:ablation}
\end{table}

\subsection{Ablation Study and Comparison with MLM}

We demonstrated the superior performance of CCBERT on
three different downstream tasks, indicating that our pre-training objectives benefit code change understanding.
In this discussion, we carry out an ablation study and understand the contribution of each pre-training objective. 
We further pre-train four model variants, each of which is trained with one pre-training objective removed, and evaluate their performance on all three downstream tasks.
The model variant CCBERT-Unit, CCBERT-New, CCBERT-Old, and CCBERT-Edit represent CCBERT trained without the
Masked Unit Prediction, the Masked New Token Prediction, the Masked Old Token Prediction, and the Masked Edit Prediction respectively.
Table~\ref{table:ablation} shows the results of the ablation study. The drop in the model performance shows the importance and contribution of each pre-training task. The results confirm that all the pre-training tasks help CCBERT to learn a better code change representation. Among them, the Masked New Token Prediction task is more important while the Masked Old Token Prediction task and the Masked Unit Prediction task are less helpful.

In addition, we are interested in comparing the effectiveness of our pre-training objectives with the well-known Masked Language Modeling (MLM) objective used in CodeBERT~\cite{CodeBERT} and GraphCodBERT~\cite{GraphCodeBERT}. CodeBERT and GraphCodBERT both employ a 12-layer Transformer encoder, which is three times larger than CCBERT. Furthermore, CCBERT is pre-trained on different pre-training data compared to CodeBERT and GraphCodeBERT. Therefore, comparing CCBERT directly with CodeBERT (GraphCodeBERT) would not reveal the advantages and drawbacks of our proposed pre-training tasks compared to the original MLM.

To provide a fair comparison between our proposed pre-training objectives and the MLM, we developed a code change pre-trained model based on MLM (namely CCMLM) that uses the same pre-training data as CCBERT. In this model, we simply concatenate the old and new versions of the code and use it as the input. We followed the default MLM pre-training objective to randomly mask 15\% of the tokens with the special token ``\texttt{<MASK>}'' and predict the masked tokens. We utilized the same pre-training data and settings as CCBERT for this model. By comparing the performance of CCBERT with this model, we can determine the effectiveness of our proposed pre-training objectives. Table~\ref{table:ablation} presents the performances of CCMLM (i.e., the Original MLM row) compared to CCBERT. Results show that CCBERT can lead to 4.9\%,  10.7\%,  and 1.2\% improvements on three tasks, respectively. We also conducted the Wilcoxon Signed Rank Test~\cite{wilcoxon1992individual} at a 95\% confidence level (i.e., $p \text{-} value < 0.05$). Results show that the improvements achieved by CCBERT are statistically significant. For instance, the p-value between CCBERT and CCMLM on task 3 is about 3e-4.

\subsection{Comparisons with Larger Pre-trained Code Models}
In this subsection, our objective is to investigate the effectiveness of CCBERT in comparison to off-the-shelf pre-trained code models that are commonly used for source code understanding, such as CodeBERT and GraphCodeBERT. Apart from effectiveness, we will also compare these models from various perspectives, which include the model size and time required for training/fine-tuning and testing~\footnote{As the speed and size of CodeBERT and GraphCodeBERT are very similar, due to limited space in the paper, we only include the comparison with CodeBERT.}. These perspectives are essential to consider for practical purposes.

\begin{table}[t]
\centering
\caption{Comparison with popular pre-trained code models in effectiveness}
\begin{tabular}{l|rl|rl|rl}
\hline
\textbf{PTM Comparison} & \multicolumn{2}{c|}{\textbf{Task1}}                                    & \multicolumn{2}{c|}{\textbf{Task2}}                                   & \multicolumn{2}{c}{\textbf{Task3}}                                   \\ \hline
\textbf{Metrics}        & \multicolumn{1}{c|}{\textbf{PR}}   & \multicolumn{1}{c|}{\textbf{ROC}} & \multicolumn{1}{c|}{\textbf{Acc.}} & \multicolumn{1}{c|}{\textbf{F1}} & \multicolumn{1}{c|}{\textbf{Acc.}} & \multicolumn{1}{c}{\textbf{F1}} \\ \hline
\textbf{CodeBERT}       & \multicolumn{1}{r|}{47.9}          & 73.5                              & \multicolumn{1}{r|}{78.8}          & 77.0                             & \multicolumn{1}{r|}{90.0}          & 90.4                            \\ \hline
\textbf{GraphCodeBERT}  & \multicolumn{1}{r|}{47.5}          & 72.8                              & \multicolumn{1}{r|}{77.1}          & 74.5                             & \multicolumn{1}{r|}{90.5}          & 90.8                            \\ \hline
\textbf{CCBERT}         & \multicolumn{1}{r|}{\textbf{50.6}} & \textbf{76.5}                     & \multicolumn{1}{r|}{\textbf{81.0}} & \textbf{80.0}                    & \multicolumn{1}{r|}{\textbf{91.6}} & \textbf{91.8}                   \\ \hline
\end{tabular}
\label{tab:PTM}
\end{table}

\begin{table}[t]
\caption{Comparison of CCBERT and other pre-trained code models in terms of the number of transformer encoder layers, trainable parameters, and GPU resources needed.}
\small
\centering
\footnotesize
\begin{tabular}{l|r|r|r}
\hline
\textbf{Models}                                                    & \multicolumn{1}{c|}{\textbf{Layers}} & \multicolumn{1}{c|}{\textbf{Param.}} &  \textbf{\begin{tabular}[c]{@{}c@{}}GPU \\ Resource\end{tabular}} \\ \hline
\textbf{CodeBERT}                                                  & 12                                   & 125M                                                                                                     & x7.9                                                             \\ \hline
\textbf{CCBERT}                                                    & 4                                    & 38M                                                                                                   & x1.0                                                             \\ \hline
\textbf{CC2Vec}                                                    &3                                     & 8M                                                                                                  &x0.6                                                              \\ \hline
\end{tabular}
\label{table:model_size}
\end{table}

\subsubsection{Effectiveness}
Table~\ref{tab:PTM} presents a comparison of the effectiveness of CodeBERT, GraphCodeBERT, and CCBERT. The results show that CCBERT consistently outperforms CodeBERT and GraphCodeBERT across the three evaluated tasks. Specifically, in JIT defect prediction (task 1), CCBERT achieves 5.6\% and 4.1\% higher AUC-PR and AUC-ROC scores than CodeBERT (6.5\% and 5.1\% higher than GraphCodeBERT). In patch correctness prediction (task 2), CCBERT leads to 2.8\% and 3.9\% (5.1\% and 7.4\%) improvements over CodeBERT (GraphCodeBERT) in terms of accuracy and F1-scores. For bug-fixing commit prediction (task 3), CCBERT achieves 1.8\% and 1.5\% (1.2\% and 1.1\%) improvements over CodeBERT (GraphCodeBERT) in terms of accuracy and F1-score. To confirm the statistical significance of these improvements, we conducted the Wilcoxon Signed Rank Test~\cite{wilcoxon1992individual} at a 95\% confidence level (i.e., $p \text{-} value < 0.05$). The test confirmed that the improvements achieved by CCBERT are statistically significant.

\begin{center}
\begin{tcolorbox}[colback={blue!5!white},
                  colframe=black,
                  width=9cm,
                  arc=1mm, auto outer arc,
                  boxrule=0.2pt,
                  top=0pt,
                  bottom=0pt
                 ]
CCBERT consistently outperforms two off-shelf pre-trained models (i.e., CodeBERT and GraphCodeBERT) on all code change-based tasks.
\end{tcolorbox}
\end{center}

\begin{table}[t]
\caption{The time cost of CCBERT, CC2Vec, and CodeBERT in both fine-tuning (training) and evaluation (testing) for three downstream tasks.}
\resizebox{\columnwidth}{!}{%
\footnotesize
\centering
\begin{tabular}{ll|r|rr|rr|rr}
\hline
\multicolumn{2}{c|}{\multirow{2}{*}{\textbf{Time Cost}}}                  & \multicolumn{1}{c|}{\multirow{2}{*}{\textbf{\begin{tabular}[c]{@{}l@{}}\# of \\ Data\end{tabular}}}} & \multicolumn{2}{c|}{\textbf{CodeBERT}}   & \multicolumn{2}{c|}{\textbf{CC2Vec}}                                   & \multicolumn{2}{c}{\textbf{CCBERT}}                                       \\ \cline{4-9} 
\multicolumn{2}{c|}{}                                                     & \multicolumn{1}{c|}{}                                 & \multicolumn{1}{c|}{\textbf{Train}} & \multicolumn{1}{c|}{\textbf{Test}} &
 \multicolumn{1}{c|}{\textbf{Train}} & \multicolumn{1}{c|}{\textbf{Test}} &
 \multicolumn{1}{c|}{\textbf{Train}} & \multicolumn{1}{c}{\textbf{Test}} \\ \hline
\multicolumn{1}{l|}{\multirow{3}{*}{\textbf{Task1}}} & \textbf{JDT}      & 4k                                                    & \multicolumn{1}{r|}{24m}             & 23s & \multicolumn{1}{r|}{3m}             &0.2s                                   & \multicolumn{1}{r|}{2m}              & 0.4s                               \\ \cline{2-9} 
\multicolumn{1}{l|}{}                                 & \textbf{Plat.} & 11k                                                   & \multicolumn{1}{r|}{75m}             & 69s   & \multicolumn{1}{r|}{10m}             &1s                                & \multicolumn{1}{r|}{7m}              & 2s                                 \\ \cline{2-9} 
\multicolumn{1}{l|}{}                                 & \textbf{Gerrit}   & 14k                                                   & \multicolumn{1}{r|}{90m}  & 91s                                                  & \multicolumn{1}{r|}{12m}            &2s                                  & \multicolumn{1}{r|}{9m}             & 2s                                 \\ \hline
\multicolumn{2}{l|}{\textbf{Task 2}}                                      & 1k                                                    & \multicolumn{1}{r|}{11m}             & 8s   & \multicolumn{1}{r|}{5s}             &0.2s                                 & \multicolumn{1}{r|}{2m}              & 1s                                 \\ \hline
\multicolumn{2}{l|}{\textbf{Task 3}}                                      & 55k                                                   & \multicolumn{1}{r|}{11.8h}            & 116s  & \multicolumn{1}{r|}{27h}            &45s                               & \multicolumn{1}{r|}{2.1h}             & 22s                                \\ \hline
\end{tabular}
}
\label{table:time}
\end{table}

\subsubsection{Model Size}
The results presented in Table~\ref{table:model_size} reveal that CCBERT has a smaller model size compared to other pre-trained models, with only 4 transformer encoder layers and 38M trainable parameters. In contrast, CodeBERT, which has 12 layers, is approximately three times larger than CCBERT, with 125M trainable parameters. Larger model sizes generally require more GPU resources for training and inference. In particular, under the same experimental conditions, CodeBERT requires approximately 7.9 times more GPU resources\footnote{To measure the GPU resources needed when fine-tuning, we experiment on a machine equipped with Intel(R) Xeon(R) CPU E5-2698 v4 @ 2.20GHz and NVIDIA Tesla V100 GPU with 16 GB of memory.} than CCBERT for both training and fine-tuning. These observations underscore the efficiency of CCBERT in terms of memory and computational resources. 

We also compare CCBERT with CC2Vec to obtain a comprehensive understanding of CCBERT's model size and GPU requirements.
Different from CCBERT and other pre-trained code models, CC2Vec is not based on Transformers but on a hierarchical attention network (mainly consisting of three layers of Gated Recurrent Units (GRUs)~\cite{GRU}) which contains less trainable parameters. CC2Vec only has 8M parameters which is much smaller than CCBERT and CodeBERT. 
In addition, CC2Vec only demands 60\% GPU resource of CCBERT which indicates that CC2Vec is more lightweight than CCBERT. 
However, CCBERT shows much better effectiveness: significantly outperforming CC2Vec in all studied tasks. Besides, CCBERT has a comparable training/inference time cost with CC2Vec (as shown in Section~\ref{time_cost}).

\begin{center}
\begin{tcolorbox}[colback={blue!5!white},
                  colframe=black,
                  width=9cm,
                  arc=1mm, auto outer arc,
                  boxrule=0.2pt,
                  top=0pt,
                  bottom=0pt
                 ]
CCBERT is about 3$\times$ smaller than CodeBERT (GraphCodeBERT) in model size and requires about 7.9$\times$ less GPU resources. 
\end{tcolorbox}
\end{center}

\subsubsection{Time Cost}
\label{time_cost}
A slow speed will impair the practical usability of the approach. We evaluate the efficiency performance of CCBERT, CC2Vec, and CodeBERT in both training/fine-tuning and testing in each task. 
Table~\ref{table:time} shows the results.
Overall, CCBERT is 6 to 10 times faster in training and is more than 5 times faster in testing as compared to CodeBERT.
Specifically, CCBERT could be trained within 2-11 minutes for Task 1 and could give all predictions within 2 seconds which is 10 and 46 times faster than CodeBERT in training and testing on average across three datasets.
When it comes to CC2Vec, specifically, CCBERT is 12.9 times and 2.0 times faster than CC2Vec in the training/inference time of the bug-fixing commit prediction (i.e., task 3) while CCBERT is much slower than CC2Vec in the patch correctness prediction (i.e., task 2). In general, CCBERT has a comparable training/inference time with CC2Vec. 

In addition, CCBERT introduces tailored designs for encoding, preprocessing, and model representation, compared to traditional token or sub-token models like CodeBERT. These modifications do not significantly affect training efficiency but require specific data preprocessing/Transformer model implementations prior to training.
To further compare the efficiency of our design and the original Transformer design, we create a Transformer model with the same model architecture parameters (e.g., the number of layers and the vocabulary size) with CCBERT. Then we finetune the Transformer model with the same training hyper-parameters (e.g., the batch size and the number of epochs) as CCBERT. For the patch correctness assessment task, CCBERT took about 2 minutes to complete the training while the created Transformer also required 1 minute 56 seconds in its training. This indicates that the tailored designs in CCBERT do not significantly affect training efficiency.

\begin{center}
\begin{tcolorbox}[colback={blue!5!white},
                  colframe=black,
                  width=9cm,
                  arc=1mm, auto outer arc,
                  boxrule=0.2pt,
                  top=0pt,
                  bottom=0pt
                 ]
CCBERT is about 6--10$\times$ and 5--30$\times$ faster than CodeBERT in finetuning and testing. 
\end{tcolorbox}
\end{center}

\begin{center}
\begin{tcolorbox}[colback=gray!10,
                  colframe=black,
                  width=9cm,
                  arc=1mm, auto outer arc,
                  boxrule=0.5pt,
                 ]
\textbf{Summary:}
Results show that CCBERT consistently outperforms larger pre-trained code models, such as CodeBERT and GraphCodeBERT, despite having a model size that is 3 times smaller. This suggests that our proposed pre-training tasks are effective in capturing the semantics of code changes and could manage to consistently outperform other much larger pre-trained code models. On the other hand, the smaller model size of CCBERT also has some benefits, including significantly reduced training time (6--10$\times$ less), inference time (5--30$\times$ less), and GPU memory usage (7.9$\times$ less).
\end{tcolorbox}
\end{center}

\subsection{Threats to Validity}

Threats to internal validity relate to errors in our experiments and implementation.
To replicate the task-specific approach for each task, we reuse the implementations released by the original works.
One threat to external validity relates to the generality of the distributed representation generated by CCBERT.
To train CCBERT, we use an existing large-scale dataset that consists of more than 663k commits (about 1M hunks after pre-processing).
However, the dataset is only based on Java programs.
In the future, we plan to construct a more diverse dataset to learn code changes in multiple programming languages.
Threats to construct validity relate to the evaluation metrics and the statistical hypothesis test that we consider. We reuse the same evaluation metrics considered in the original downstream tasks.
We use a standard statistical hypothesis test, Wilcoxon Signed Rank Test~\cite{wilcoxon1992individual}, to check whether the performance difference between two competing approaches is significant. This test has been used in many past studies, e.g.,~\cite{xu2021post2vec,rajbahadur2019impact,wang2018deep}.
Most of the metrics and the statistical hypothesis test are well known. Thus, we believe that this threat is minimal.

\section{Related Work}
\label{sec:related}

Most of the work related to code changes are targeted for a specific task like just-in-time defect prediction. PatchNet~\cite{patchnet} and DeepJIT~\cite{DeepJIT} utilized CNN-based neural networks to extract the representation from code change hunks and used the representations to conduct bug-fix commit prediction and just-in-time defect prediction respectively. 
Commit2Vec~\cite{cabrera2021commit2vec} applied code representation learned by Code2vec~\cite{alon2019code2vec} and another pretext task of predicting the priority of Jira tickets. They evaluated their code representations in the binary classification of security-related commits. 
Dong et al.~\cite{Dong2022FIRAFG} proposed a commit message generation approach that turns code changes into graphs. They turned code change hunks into chopped ASTs, extracted edit information by comparing chopped ASTs, and formed a graph by considering both the token information and the edit information. Then, they adopted a graph-neural-network-based encoder to embed the code change graphs and use a Transformer decoder to generate the commit message.

Another related work is proposed by Yin et al.~\cite{yin2019ICLR} which represents the salient information of an edit.
Their method combined both syntactic and semantic information of code edits to learn the representation of edits and evaluated the effectiveness of their learned representation in the task of generating the new code given the old code.
The approach proposed by Yin et al. also produces code change representation. The major differences between their work and ours are: 1) Different goals. CCBERT aims to learn a generalizable representation that can be used in diverse downstream tasks. However, Yin et al.’s work aims to get good performance specifically for the source code edit task. 2) Different supervision signals. Their work generates the new version of code from the old version of code. Thus, they use the new version of code (a whole sequence) as the supervision signal. On the other hand, CCBERT separately predicts each masked code token and each masked edit action which are fine-grained token-level supervision signals. 3) Different Frameworks. Their work uses the auto-encoder framework with LSTM~\cite{lstm} and GNN~\cite{ggnn} to learn the representation of changes while CCBERT uses the Transformer encoder and follows BERT~\cite{bert} which is a general framework leveraging a large corpus of unlabeled data.

In summary, the above-mentioned works do not aim for a generalizable representation for multiple tasks but mainly aim to produce a decent performance on a specific task. Differently, CCBERT aims to learn a generalizable representation that can be used in multiple downstream tasks by using large-scale pre-training data and novel pre-training objectives.

\section{Conclusion}
\label{sec:conclusion}

In this work, we propose CCBERT, a code change representation model via self-supervised pre-training. 
CCBERT is pre-trained on a large-scale unlabeled code change dataset with four novel pre-training objectives designed for code changes.
We evaluate CCBERT on three downstream code tasks (just-in-time defect prediction, patch correctness prediction, and bug-fixing commit prediction).
Experimental results show that CCBERT significantly outperforms the prior state-of-art code change representation model CC2Vec or task-specific state-of-the-art approaches by 7.7\%--14.0\% in terms of different metrics.
We further compare the effectiveness of CCBERT with other larger pre-trained code models such as CodeBERT. Experimental results demonstrate that CCBERT achieves consistently better results than other large pre-trained code models with 7.9 times less GPU memory, 6--10 times faster in training, and 5--30 times faster in inference.
In our future work, we plan to leverage commit messages to improve CCBERT by filtering out low-quality commit messages.
We are also interested in applying CCBERT to other code change-related tasks.
A replication package of our work is available at \url{https://github.com/soarsmu/CCBERT}.

\vspace{0.1cm}
\noindent{\bf Acknowledgement.} This research / project is supported by the National Research Foundation, Singapore, under its Industry Alignment Fund – Pre-positioning (IAF-PP) Funding Initiative. Any opinions, findings and conclusions or recommendations expressed in this material are those of the author(s) and do not reflect the views of National Research Foundation, Singapore.

\balance
\bibliographystyle{IEEEtran}
\bibliography{reference}
\balance

\end{document}